\documentclass[showpacs,twocolumn,preprintnumbers,amsmath,amssymb,floatfix,10pt]{revtex4}
\usepackage{epsfig}
\usepackage{bm}
\usepackage{graphicx}
\begin{document}

\title{Orbital magnetic moments in pure and doped carbon nanotubes}
\author{Magdalena Marga\'{n}ska\footnote{corresponding author: magda@phys.us.edu.pl}, Marek Szopa, El\.{z}bieta Zipper}
\affiliation{Department of Theoretical Physics, University of Silesia, ul Uniwersytecka 4,
40 007 Katowice, Poland}

\begin{abstract}
The unusual band structure of carbon nanotubes (CNs) results in their remarkable
magnetic properties. The application of magnetic field parallel to the tube axis can change 
the conducting properties of the CN from metallic to semiconducting and vice versa.
Apart from that {\bf B} induces (via the Bohm-Aharonov effect) orbital magnetic moments $\mu_{orb}$
in the nanotube. These moments are studied both in pure
and hole- or electron-doped CNs, isolated or in a circuit. Remarkably, $\mu_{orb}$ in pure
CNs depends uniquely on their original conducting properties, length, and temperature but
it does not depend on the nanotube radius or the particular chirality. In doped nanotubes
the magnetic moments can be strongly altered and depend on the radius and chirality. 
Temperature can even change their character from diamagnetic at low $T$ to paramagnetic at 
high $T$. A full electron-hole symmetry in doped tubes is also revealed.
\end{abstract}



\maketitle


\section{Introduction}
The electronic fate of a carbon nanotube (CN) is in general determined once it has been grown.
Depending on the radius and the chiral angle it can be semiconducting or metallic.
The electronic properties of CN can however be modulated by coaxial magnetic field
via the Bohm-Aharonov (BA) effect which can turn a metallic CN into a semiconducting
one and vice versa. This has been predicted by Ajiki and Ando \cite{ajiki1} and observed 
recently in three independent measurements \cite{mceuen,bezryad,smalley}.
CN have an ideal structure for studying the effect of the BA flux on the energy spectrum.
A magnetic field {\bf B} introduces a phase factor in the electron wave function in the
circumferential direction and leads to a shift in the energy bands. 
The value of the energy shift depends on the strength of the applied field and on the
orbital magnetic moment. This can be observed 
e.g. as a change in the band gap structure by measuring the conductance of a single
nanotube suspended between two electrodes  \cite{mceuen,bezryad}. The magnetic field 
influences the motion of electrons around the circumference of CN, giving rise to 
persistent currents \cite{buttiker,gefen} which at low temperatures do not decay. Persistent current multiplied 
by the CN cross section give the
orbital magnetic moment $\mu_{orb}$ directed along the axis.\\
In the present paper we calculate $\mu_{orb}$ in single-wall CNs in the tight binding 
approximation for various
chiralities, lengths and radii (provided that $R > 10$\AA), for a range of electron or hole doping
values. We study both isolated 
nanotubes, where the number of electrons $N_e = const$ and nanotubes connected to
a particle reservoir (e.g. as a part of a circuit), where the chemical 
potential $\mu_{chem}=const$.\\ 
We find that  in undoped nanotubes the character of the magnetic moment depends
only on the CN's conducting properties (i.e. whether the tube is metallic or semiconducting),
its length and the temperature of the system.
On the other hand, in hole- or electron doped nanotubes the behaviour of $\mu_{orb}(\phi)$ depends strongly
on the chirality of the nanotube, on  its size, and on the degree of hole or electron doping.\\
In our model calculations we study nanotubes in the ballistic regime and we work in
the non-interacting electrons approximation which has yielded good agreement with
experimental results in mesoscopic rings \cite{mailly} and in carbon nanotubes \cite{bezryad,stm,roche}.


\section{The model}

Carbon nanotubes are commonly considered and analysed as rolled-up 
graphene planes. We follow here the same approach, working in the basis
in which the lattice generators are $\mathbf{T}_1 = \sqrt{3} e_x$, 
$\mathbf{T}_2 = \sqrt{3}/2 e_x + 3/2 e_y$, and the length unit is the length of the $C-C$ bond, 1.42{\AA} \cite{gonz,book}. 
A nanotube is  uniquely defined by four parameters $(m_1, m_2)\times(p_1, p_2)$, 
which define its circumference and length vectors:
\begin{eqnarray}
\mathbf{L}_{\perp} & = & m_1 \mathbf{T}_1 +  m_2 \mathbf{T}_2 \hspace{0.5cm} \textnormal{for the circumference,} \nonumber \\
\mathbf{L}_{\parallel} & = & p_1 \mathbf{T}_1 +  p_2 \mathbf{T}_2 \hspace{0.5cm} \textnormal{for the length.} \label{eq:lenghts}
\end{eqnarray}
In most theoretical works on carbon nanotubes the tube under examination is
considered to be infinitely long, resulting in a continuous spectrum of momenta
along the CN. Since the real CNs are very long but finite, we consider $L_{\parallel} < \infty$. 
Therefore, the momentum is quantized in both directions.  
We chose the longitudinal boundary 
conditions to be cyclic:
\begin{equation}\label{boundary-l}
\mathbf{k}\cdot\mathbf{L}_{\parallel} = 2\pi \l_{\parallel}, \hspace{1cm} l_{\parallel} \in \mathbb{Z},
\end{equation}
where  $\parallel$ stands for the direction parallel to the length of the nanotube.
We found no significant differences between currents in cyclic and open
longitudinal boundary conditions (see also \cite{bloch}).\\
The magnetic field is applied parallel to the axis of the CN. The Aharonov-Bohm phase 
factor modifies the transverse boundary condition (in the direction
perpendicular to the magnetic field):
\begin{equation}\label{ahbohm}
\mathbf{k}\cdot\mathbf{L}_{\perp} = 2\pi (\l_{\perp} + \frac{\phi}{\phi_0}), \hspace{1cm} l_{\perp} \in \mathbb{Z},
\end{equation}
where $\perp$ stands for the direction parallel to the circumference of the nanotube,
$\phi$ is the magnetic flux and $\phi_0 = h/e$ is the quantum flux unit. The 
momentum vector $\mathbf{k}$ can be defined either in the $(k_x, k_y)$ or $(k_\perp, k_\parallel)$
basis. We use the second one throughout our calculations, the first only in the dispersion relation.
When $\phi \neq 0$, the currents carried by states with $\mathbf{k}$ and $-\mathbf{k}$
do not cancel out, and a net current appears. At fields accessible
in labs only a part of the Bohm-Aharonov period $\phi_0$ could be observed in 
nanotubes of small diameter. For example, in a CN with radius $R = 25${\AA} a magnetic field $B$
of the order of 210T is required to observe the full
$\phi_0$ period.\\
Currents running along the circumference of the CN induce in the nanotube a 
magnetic moment parallel to its axis. The magnetic moment of an electron close
to the Fermi level can also be calculated by the following reasoning.\\
The energy gap between conduction and valence states close to the Fermi surface (FS) is
\begin{equation}
E_g^0 = \hbar v_F (k_{\perp} - K_i),\hspace{0.5cm} i = 1,2, 
\end{equation}
where $K_i$ are the Fermi points where the valence and conduction bands of graphene meet.
The Bohm-Aharonov effect shifts the allowed $k_{\perp}$ by $\delta  k_{\perp}(\phi)$, 
\begin{equation}
k_{\perp}(\phi) = k_{\perp} + \delta  k_{\perp}(\phi), \hspace{0.5cm} \delta  k_{\perp}(\phi) = \frac{\phi}{R\phi_0},
\end{equation}
resulting also in the energy shift $\Delta E$ \cite{mceuen},
\begin{equation}
\Delta E \simeq \frac{\partial E}{\partial k_{\perp}}\mid_{k_F} \delta  k_{\perp}(\phi) = \pm \frac{\hbar v_F}{R} \frac{\phi}{\phi_0} = \overrightarrow{\mu}^F_{orb}\cdot \mathbf{B},
\end{equation}
where $\overrightarrow{\mu}^F_{orb}$ is the orbital magnetic moment of an electron 
at (or the closest to) the Fermi surface
\begin{equation}
\overrightarrow{\mu}^F_{orb} = \frac{R e v_F}{2} {\mathbf e}_{\parallel}
\end{equation}
(${\mathbf e}_{\parallel}$ is the unit vector along the CN axis). 
This shift of the energy states results in the change of the band gap and can convert a metallic
CN into a semiconducting one and vice versa. Therefore, from now on by 'metallic' and 
'semiconducting' we understand CN's which display this behaviour
at $\phi = 0$.\\
The change of the band gap with the magnetic field has been investigated  in three independent
measurements \cite{mceuen,bezryad,smalley} and in the first one helped to determine the magnitude of the orbital magnetic moment
of an electron at the FS, $\mu^F_{orb}$. A good agreement between the experimental and
the theoretical values has been found \cite{mceuen}. Nevertheless,  the magnetic response of a CN is determined by its full magnetic moment, given by
\begin{equation}
\mu_{orb}(\phi, T) = -\pi R^2 \frac{\partial F(\phi, T)}{\partial\phi},
\end{equation}
where $F(\phi, T)$ is the free energy of the CN. In other words, $\mu_{orb}$ can be written
in terms of the total current $I(\phi, T)$, which runs at the cylindrical surface in the presence
of the magnetic field (for detailed derivation see \cite{actapol}).
\begin{eqnarray}\label{muorb-with-T}
& \mu_{orb}(\phi, T)  =  \pi R^2 I(\phi, T) = & \\
  = & \pi R^2 \sum_{k_\perp, k_\parallel} \frac{1}{1 + \exp\left[(E_{\mathbf{k}}(\phi) - \mu_{chem}(\phi))/kT\right]}I_{\mathbf{k}}(\phi). &  \nonumber
\end{eqnarray}
The sum runs over the whole Brillouin zone (BZ).
This current is persistent at low $T$. In our paper we investigate the dependence 
of $\mu_{orb}(\phi, T)$ on temperature, nanotube parameters (radius, chirality, length) 
and the value of the chemical potential which changes with doping.
This is an extension of the work of Ajiki and Ando \cite{ajiki1}, who calculated the magnetic moment
$\mu_{orb}$ for an undoped tube at $T = 0$.\\

The currents carried by individual momentum states at $T = 0$ are
\begin{equation}\label{length-to-width}
I_{\mathbf{k}}(\phi) = - \frac{\partial E_{\mathbf{k}}}{\partial\phi} = 
	 - \frac{\partial E_{\mathbf{k}}}{\partial k_{\perp}}\;\frac{\partial k_{\perp}}{\partial\phi}
 =  - \frac{\partial E_{\mathbf{k}}}{\partial k_{\perp}}\;\frac{2\pi}{\phi_0 |\mathbf{L}_{\perp}|},	
\end{equation}
where the form of $E(\mathbf{k})$ depends on the approximation used. Our choice 
is the tight binding approximation, commonly used in calculations involving graphene
structures. Consequently, the dispersion relation is \cite{gonz,book}
\begin{equation}\label{dispersion}
E_{\mathbf{k}} = \pm \gamma \sqrt{1 + 4\cos^2(\frac{\sqrt{3}}{2}k_x)
+ 4\cos(\frac{\sqrt{3}}{2}k_x)\cos(\frac{3}{2}k_y)},
\end{equation}
where $\gamma$ is the hopping integral for graphene. Various authors assume different
values for $\gamma$, ranging from 2.5 eV to  3 eV \cite{bezryad,chuu}. \\ 
The individual currents $I_{\mathbf{k}}(\phi)$ in Eq. (\ref{length-to-width}) depend on
the position of the {\bf k} states in the Brillouin zone. The level of $\mu_{chem}$ defines the Fermi surface
and determines
the range of states whose contributions dominate the sum (\ref{muorb-with-T}). 
The distribution of allowed momentum states in the BZ depends on the chirality of the CN.
Therefore at shifted $\mu_{chem}$ different states fall into the dominating range 
depending on the chirality of the nanotube. Consequently the total magnetic moment 
depends on the doping level and on the chirality of the CN.\\
At the half-filling, corresponding to undoped nanotube, the chemical potential
is 0 ($\mu_{chem} = 0$). Doped systems (either doped with electrons or with holes) can be 
studied under two
different physical conditions.\\
 If the system is connected with a particle reservoir, its chemical potential at a given doping is constant ($\mu_{chem} = const$).  The
distribution of allowed momentum states in the BZ shifts with the magnetic flux which induces
a shift of the Fermi level. But the chemical potential is determined
externally and the CN can absorb the number of electrons or holes
necessary to keep it stable. The number of electrons then changes with the 
magnetic flux, $N_e = N_e(\phi)$. \\
When the nanotube is isolated, there is no exchange of electrons with a reservoir, their number is
constant ($N_e = const$) and the chemical potential is a function of the magnetic flux 
($\mu_{chem} = \mu_{chem}(\phi)$). The magnetic moments have in general different shape
in isolated than in connected nanotubes. We performed the calculations in both cases.\\
We have calculated numerically the magnetic moment according to Eq. (\ref{muorb-with-T}). 
In the case of constant number of electrons the chemical potential has been found
at each value of $\phi$ from the condition $\sum_{(k_\perp, k_\parallel)} f_{FD}(E_{(k_\perp, k_\parallel)}( \phi)) = N_e$, 
where $f_{FD}(E_{\mathbf{k}}(\phi))$ is the Fermi-Dirac distribution function. 


\section{Orbital magnetic moments -- undoped nanotubes}

Several effects concerning the persistent currents and induced magnetic moments
in parallel magnetic field can be deduced from an analysis of the structure of the energy
spectrum and the Brillouin zone. In this Section we discuss undoped nanotubes
i.e. $\mu_{chem} = 0$.
In this case there is no difference between the $\mu_{chem} = const$ and $N_e = const$
conditions. The number of states below the Fermi level is the same
as in the Brillouin zone, consequently, $N_e = const$ regardless of the value of the external
magnetic flux. The energy gap opens or closes with changing $\phi$, but the chemical potential
is fixed at 0.\\  

\begin{figure}
\includegraphics[width=7cm]{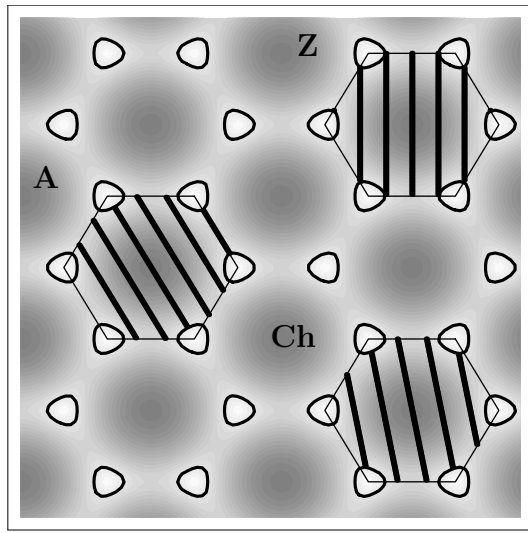}\\
\begin{picture}(0,0)
\put(-90,140){\bf \large A}
\put(10,190){\bf \large Z}
\put(0,80){\bf \large Ch}
\end{picture}
\caption{\label{fig:plane-holes} The reciprocal lattice of graphene, with 
momentum states and  first Brillouin zones of {\bf Z}igzag (5,0), 
{\bf A}rmchair (3,3), and {\bf Ch}iral (4,1) nanotubes doped to $-0.6\gamma$. 
The Fermi contours are the thick triangular loops and the thick straight lines correspond
to $k_\perp = const$ lines.
In udoped nanotubes all momentum states in the Brillouin zone enter into the 
sum from Eq. (\ref{muorb-with-T}) and $\mu_{orb}(\phi)$ is chirality-independent. 
In doped nanotubes only those states which lie below the 
Fermi level (outside the loops) contribute to the sum. Note that the number of missing states 
(the missing fragments of $k_\perp = const$ lines) is different in {\bf A}, {\bf Z} and {\bf Ch} cases, 
which results in the chirality dependence of $\mu_{orb}(\phi)$.}
\end{figure}

\begin{figure}
\includegraphics[width=6cm]{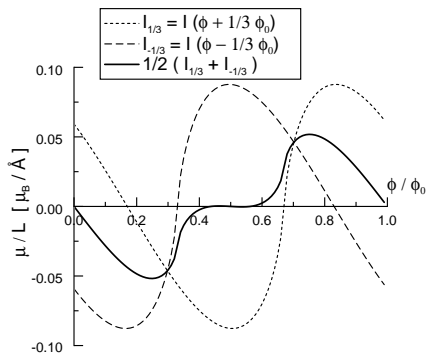}\\
\caption{\label{fig:whysemicond} The magnetic moment per {\AA} in a semiconducting nanotube
as a superposition of two metallic moments shifted by $\pm \phi_0/3$.}
\end{figure}

{\bf $\mu_{orb}(\phi, T)$ in metallic and semiconducting CN's.}
There are two types of behaviour of $\mu_{orb}(\phi, T)$. In metallic nanotubes
it is paramagnetic at $\phi = 0$, in semiconducting ones it is diamagnetic. 
This effect does not depend on the particular chirality of the metallic or semiconducting 
nanotube. 
%
%
%
\begin{figure}
\includegraphics[width=6cm]{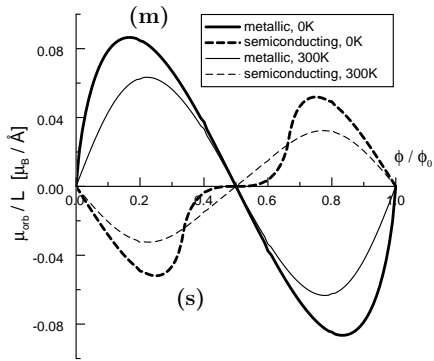}\\
\caption{\label{fig:chirality-independent} Orbital magnetic moment per {\AA} in a single-wall
nanotube of radius 25\AA, for different chiralities. All metallic nanotubes 
have the same (up to 1\%) $\mu_{orb}(\phi)$. The same is true for the semiconducting nanotubes.
Thick lines are orbital magnetic moments at T = 0K, thin lines -- at T = 300K. }
\end{figure}
%
%
It is so because the main component of the magnetic moment (\ref{muorb-with-T}) comes
from the $\mathbf{k}$ states close to the Fermi points $\mathbf{K}_i$. The dispersion relation
in the neighbourhood of these points has a form of two cones and therefore rotational symmetry
\cite{pseudospins}.
Because of that $\partial E/\partial k_\perp$ is independent of the particular angle (determined by
$m_1$ and $m_2$) at which $k_\perp = const$ lines (see Fig. \ref{fig:plane-holes}) lie with respect to the edges of the Brillouin zone.
%
%
%
%
%
%
In the case of metallic CNs these momentum lines cross the Fermi points (see  Fig \ref{fig:bzones}a)
and the magnetic moment has the $(m)$ shape from Fig. \ref{fig:chirality-independent}. 
Due to the peculiar band structure of the CNs,
the magnetic moment in a semiconducting CN ($s$) has an unusual shape, with a plateau
around $\phi_0/2$.
Close inspection of the structure of momentum lines in the BZ shows that 
it is a superposition of two metallic
moments, shifted by $\pm  \phi_0/3$ (it is shown in Fig. \ref{fig:whysemicond}). They
are generated by the two momentum ($k_\perp =  const$) lines which reach the Fermi point, one at
$\phi = \phi_0/3$, the other at $2\phi_0/3$, independently of chirality. The amplitude of 
this sum is then smaller
than in the metallic CN, where both momentum lines reach the Fermi points at the same
$\phi$. The slope of $\mu_{orb}$ vs. $\phi$ (see Fig \ref{fig:chirality-independent}) is steeper for these values of
$\phi$ where the CN is metallic, since the paramagnetic behaviour of the $\mu_{orb}$
is caused by momentum states crossing the Fermi surface, whose contributions to the
magnetic moment are the most significant. On the contrary, the gentle
slope of $\mu_{orb}$ in the semiconducting regime is caused by the diamagnetism of
the states below the FS.\\
%
%
%
%
%
{\bf Temperature dependence.} 
The magnitude of the orbital magnetic moment, contrary to the spin magnetic moment,
decreases with temperature although this decrease is not very dramatic  for the ranges 
of magnetic field accessible in labs. It is more important for CN
with larger R, i.e. for multiwall CN, and the measurements should then be performed at
lower $T$.
Since the main effect of temperature on $\mu_{orb}(\phi, T)$ on undoped nanotubes is a 
suppression of its amplitude (cf. Fig. \ref{fig:chirality-independent}), we shall in this Section 
assume $T = 0$ and analyse only $\mu_{orb}(\phi, 0)$ which we shall denote by $\mu_{orb}(\phi)$.\\

\begin{figure}
\includegraphics[width=7cm]{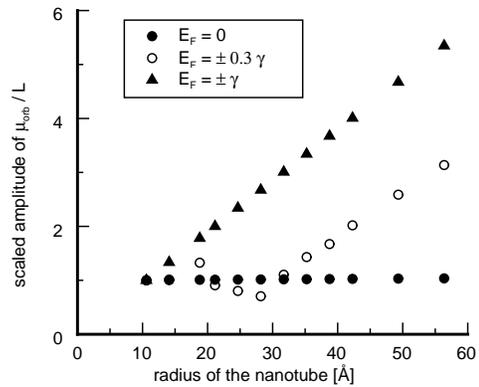}\\
\caption{\label{fig:irsquare} The dependence of the magnetic moment per unit length
on the nanotube radius in zigzag nanotubes ranging from $(27,0)$ to $(144,0)$, at half-filling and
at two values of doping. The amplitude of the $\mu_{orb}$ on the vertical axis is scaled 
with respect to the values obtained for the first nanotube, with $R = 10.59$\AA.}
\end{figure}

When the distance between momentum lines $\Delta k_\perp$
becomes small enough (i.e. when  $R > 10$\AA), $I_{\mathbf k} (\phi, T)$ is linear in $\Delta k_\perp$ 
and two effects appear. \\
{\bf Length-scaling of the magnetic orbital moments.} The sum of  $I(\mathbf{k})$ grows linearly 
with the number of states on one momentum ($k_\perp = const$) line,
which is proportional to the length of the nanotube. This is also true for doped
nanotubes.\\
{\bf Independence of $\mu_{orb} (\phi)$ of the nanotube radius.} The 
radius of the nanotube affects  $\mu_{orb} (\phi)$ in two ways.\\
1) The cross section of a nanotube $\sim R^2$.\\
2) From Eq. (\ref{length-to-width}) we conclude that the current of an individual
$\mathbf{k}$ state  $\sim 1/R$ because $\partial E_{\mathbf{k}}/\partial k_\perp$ is
constant for a given $\mathbf{k}$. The summation over the whole Brillouin zone
in Eq. (\ref{muorb-with-T}) yields another $1/R$ factor and as a result the whole sum
is proportional to $1/R^2$.\\
These two effects lead to $\mu_{orb} (\phi)$ nearly independent of $R$ (see black circles in 
Fig. \ref{fig:irsquare}), in agreement with \cite{ajiki1}.  
This result in undoped CN's does not depend on the chirality of the nanotube.
Note that $\mu^F_{orb}(\phi)$ calculated and 
measured in \cite{mceuen} increases linearly with $R$ because it is due only to electrons
closest to the Fermi level.\\

Thus for any undoped metallic or semiconducting 
nanotube the total magnetic moment depends only on its length and on temperature, and has either
the $(m)$ (in metallic CN's) or the $(s)$ (in semiconducting CN's) form from 
Fig. \ref{fig:chirality-independent}.\\
%
%
%
%
%
%
%
%
\section{Orbital magnetic moments -- doped nanotubes}

\begin{figure*}
\includegraphics[width=14cm]{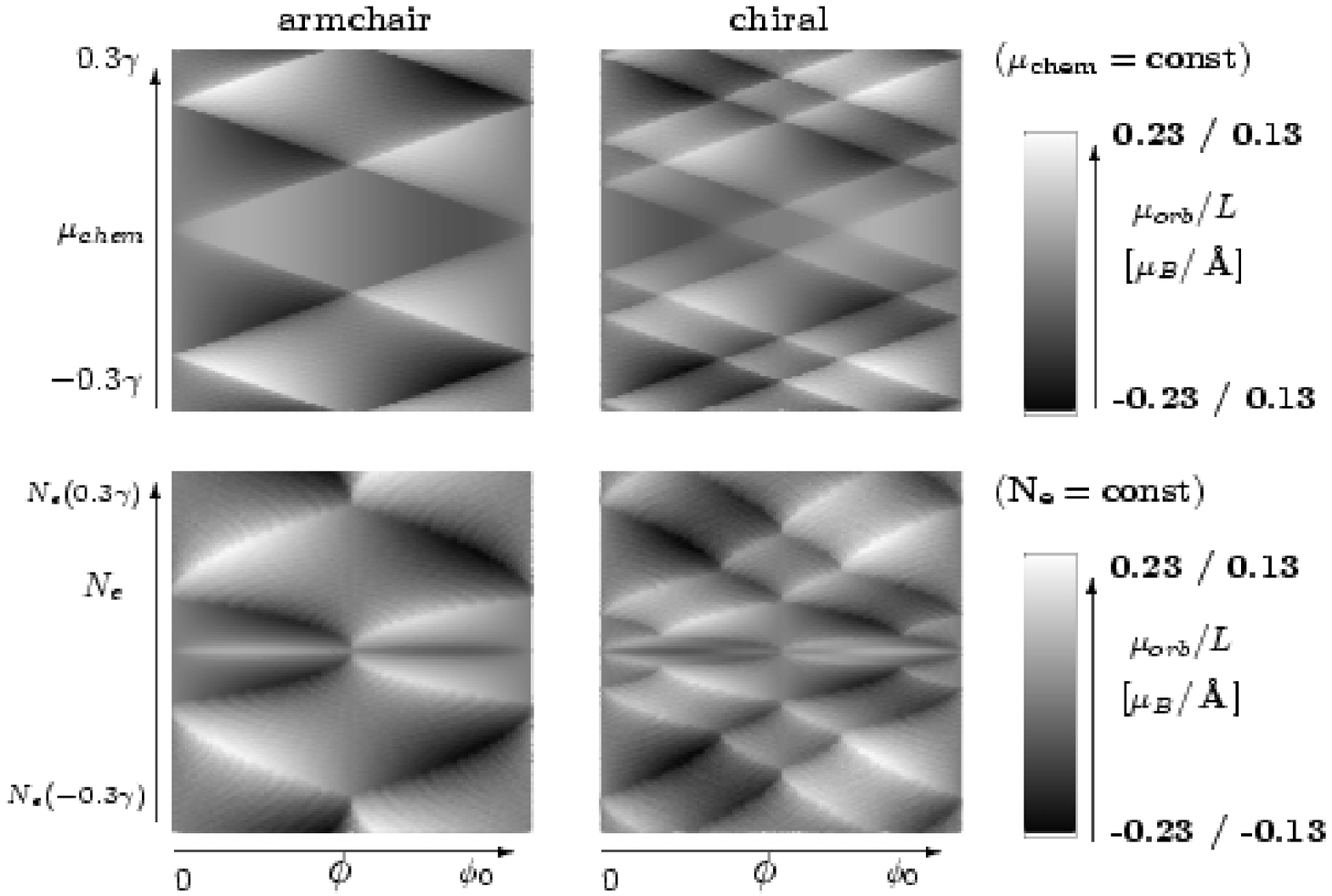}\\
\caption{\label{fig:smallcone-both} The comparison between (top) $\mu_{chem} = const$ and 
(bottom) $N_e = const$ conditions on the conical part of the dispersion relation, i.e. for $|\mu_{chem}| \leq 0.3\gamma$. Left column shows the orbital magnetic moment per {\AA} (colour scale) in 
armchair (15,15),  right column - in chiral semiconducting (15,14) nanotubes.}
\end{figure*}

\begin{figure}
\includegraphics[width=6cm]{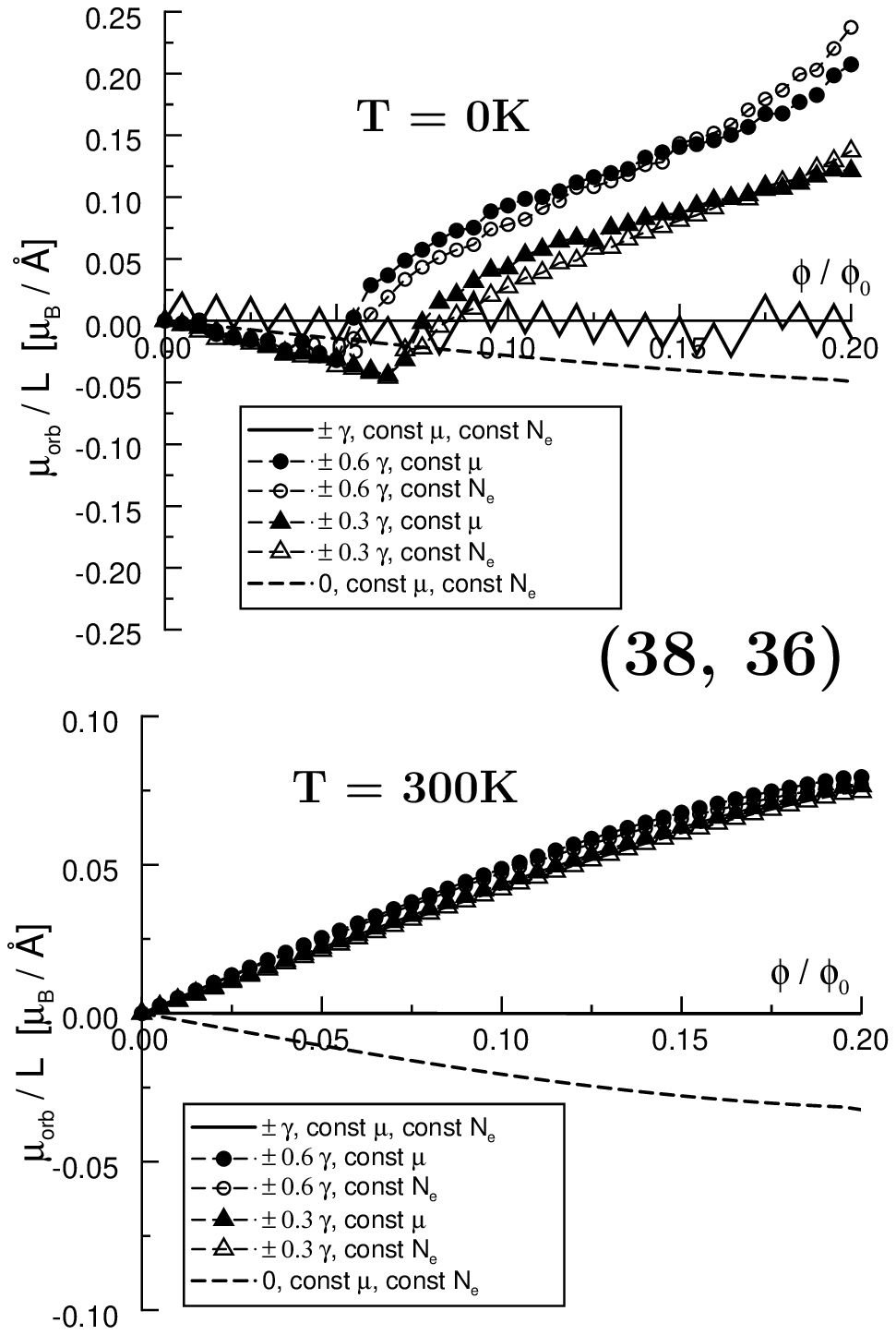}\\
\caption{\label{fig:armch} The magnetic moment per unit length in a nearly armchair 
$(38,36)\times(-4028,4101)$ nanotube, at  $E_F = 0, \pm0.3 \gamma, \pm0.6 \gamma, \pm \gamma$, both for  $\mu_{chem} = const$ and $N_e = const$, at $T = 0$K and $T = 300$K. 
$R=25$\AA, $\phi = 0.2\phi_0$ corresponds to $B = 40$T. The solid line, if not visible,
coincides with the horizontal axis. At 300K the results
for $\mu_{chem} = const$ and $N_e = const$ almost overlap.}
\end{figure}

\begin{figure}
\includegraphics[width=6cm]{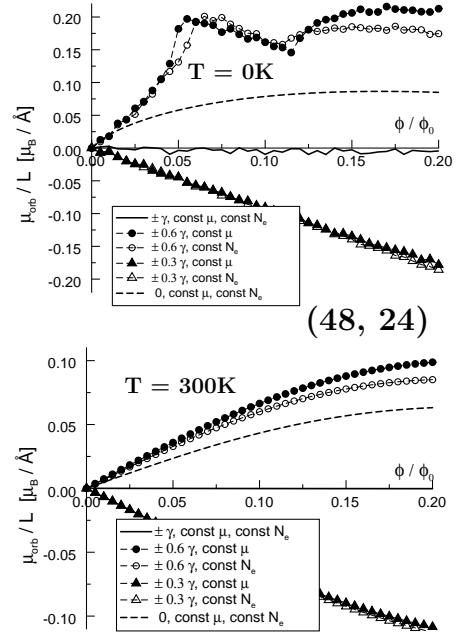}\\
\caption{\label{fig:metal} The magnetic moment per unit length in a chiral metallic (48,24)  
nanotube, at  $E_F = 0, \pm0.3 \gamma, \pm0.6 \gamma, \pm \gamma$, both for $\mu_{chem} = const$ and $N_e = const$, at $T = 0$K and $T=300$K. 
$R=25$\AA, $\phi = 0.2\phi_0$ corresponds to $B = 40$T. The solid line, if not visible,
coincides with the horizontal axis.}
\end{figure}

\begin{figure}
\includegraphics[width=6cm]{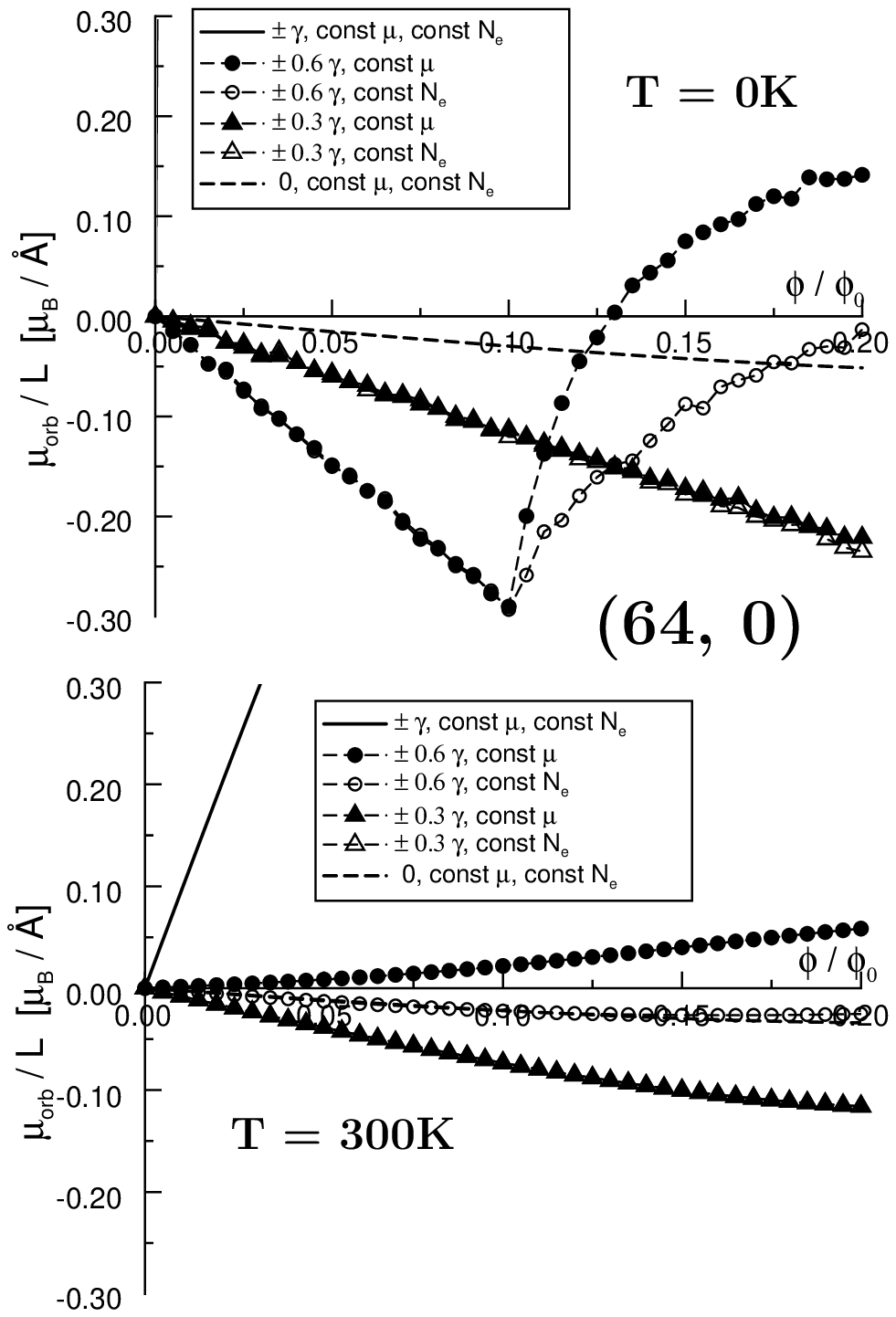}\\
\caption{\label{fig:zigzag} The magnetic moment per unit length in a zigzag (64,0)  
nanotube, at  $E_F = 0, \pm0.3 \gamma, \pm0.6 \gamma, \pm \gamma$, both for $\mu_{chem} = const$ and $N_e = const$, at $T = 0$K and $T=300$K. 
$R=25$\AA,  $\phi = 0.2\phi_0$ corresponds to $B = 40$T. The magnetic moment
at $\mu_{chem} = \pm\gamma$ and $T = 0$K in the zigzag nanotube is a very steep function
of $\phi$ (see \cite{PhyLe}) and overlaps with the vertical axis on this plot. At 300K the results
for $\mu_{chem} = const$ and $N_e = const$ overlap.}
\end{figure}

\begin{figure*}
\includegraphics[width=14cm]{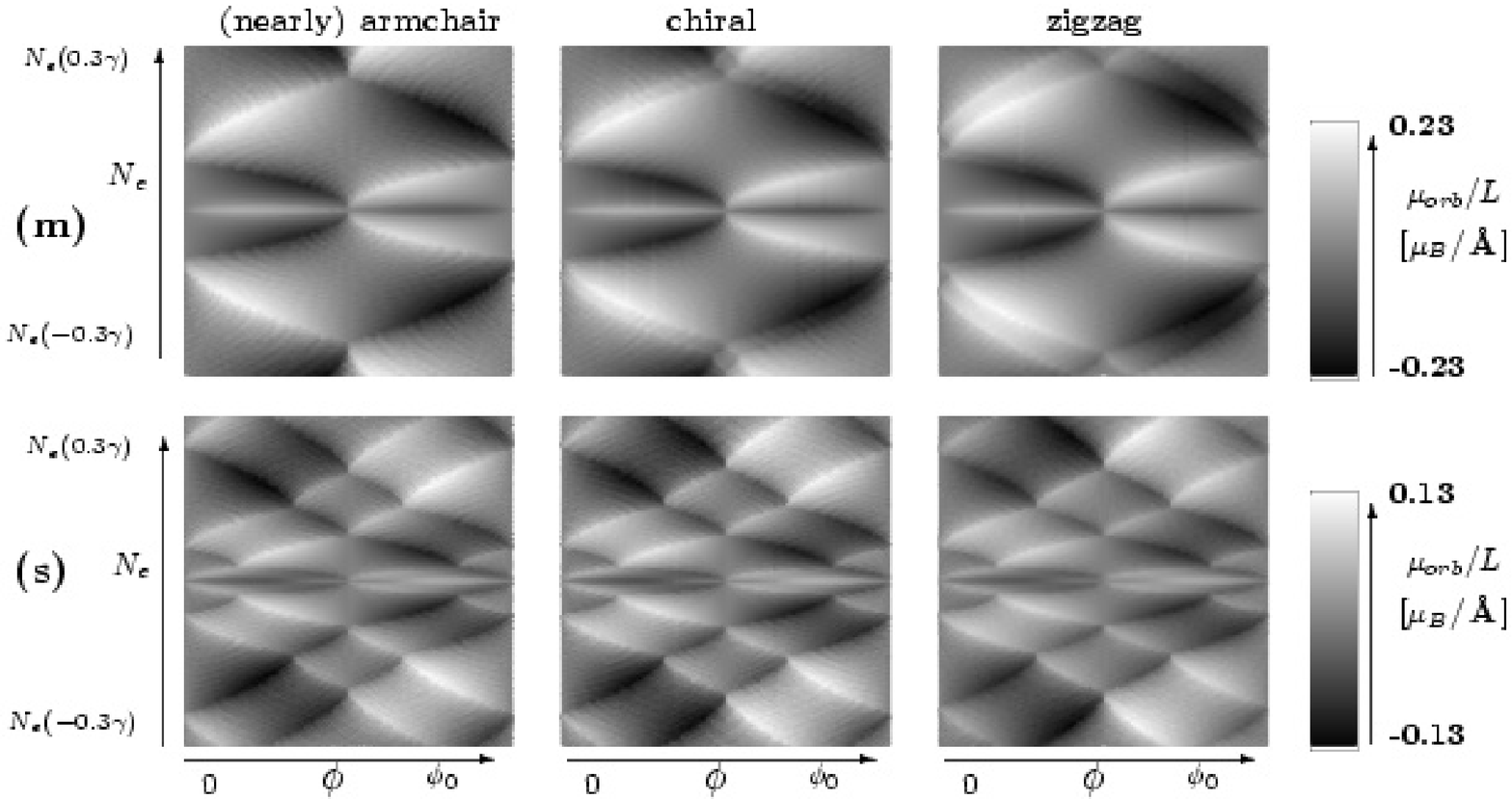}\\
\caption{\label{fig:smallcone-constN} The orbital magnetic moment per {\AA} (colour scale) for $|\mu_{chem}| \leq 0.3\gamma$,
for metallic (top row) armchair (15,15), chiral (19,10), zigzag (24,0) nanotubes, and for 
semiconducting (bottom row) nanotubes with similar chiralities -- (15,14), (19,9) and (25,0). $R = 10.2$\AA, $T = 0$. The Fermi contour around a Fermi point in this regime
is almost circular, which accounts for the similarities between the three cases.}
\end{figure*}

The peculiar band structure of graphene is reflected in the variety of behaviours of CNs 
with different chiralities upon doping. We performed the calculations of the
full magnetic moment $\mu_{orb}(\phi,T)$ for 
zigzag, armchair and chiral nanotubes. We found that it depends significantly on doping; 
being the sum of the terms
below the FS it depends strongly on the shape of the FS which changes with the number of holes or electrons introduced in the system.\\
%
%
%
\begin{figure}
\includegraphics[width=4cm]{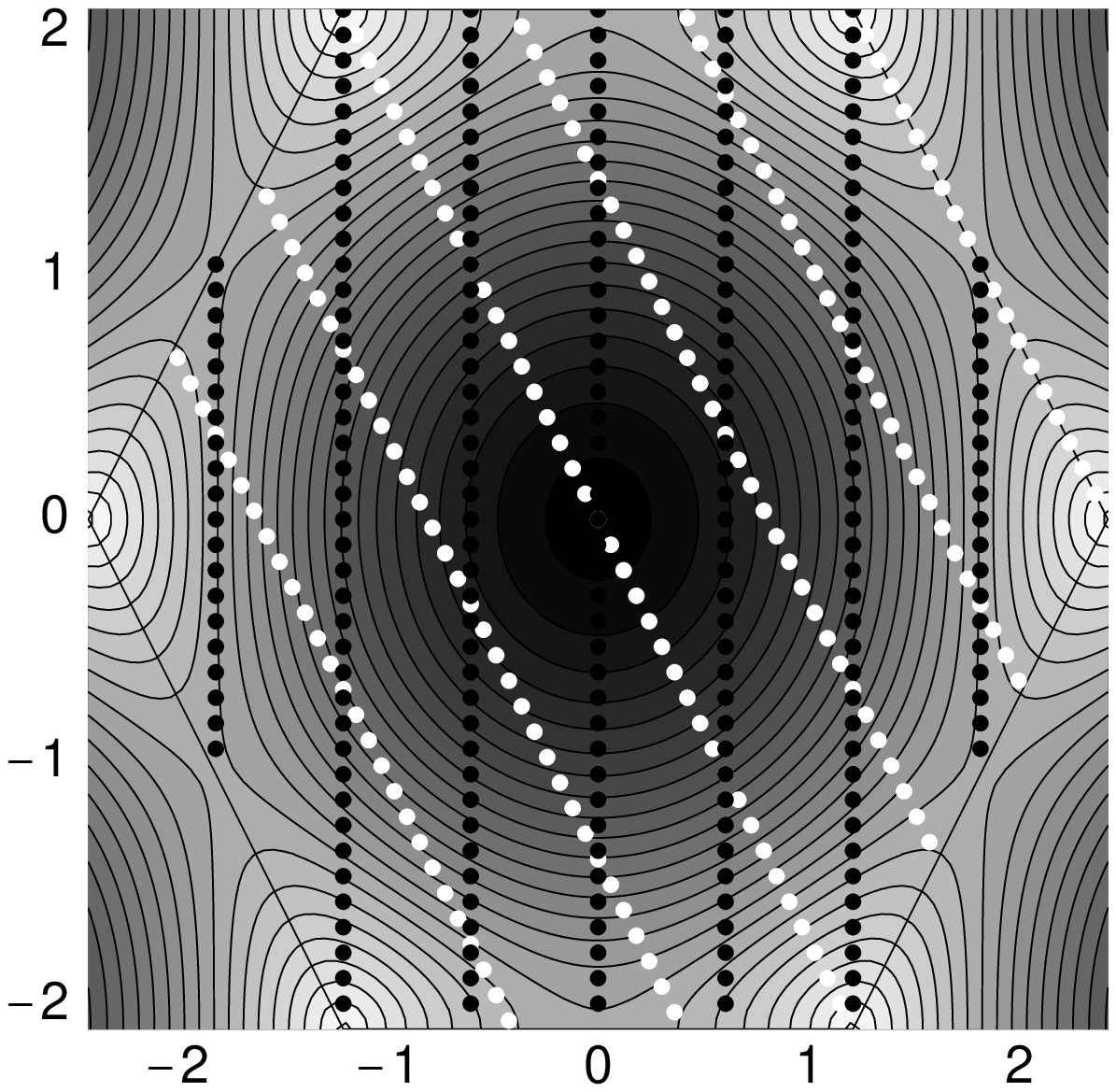}\hspace{0.5cm}
\includegraphics[width=4cm]{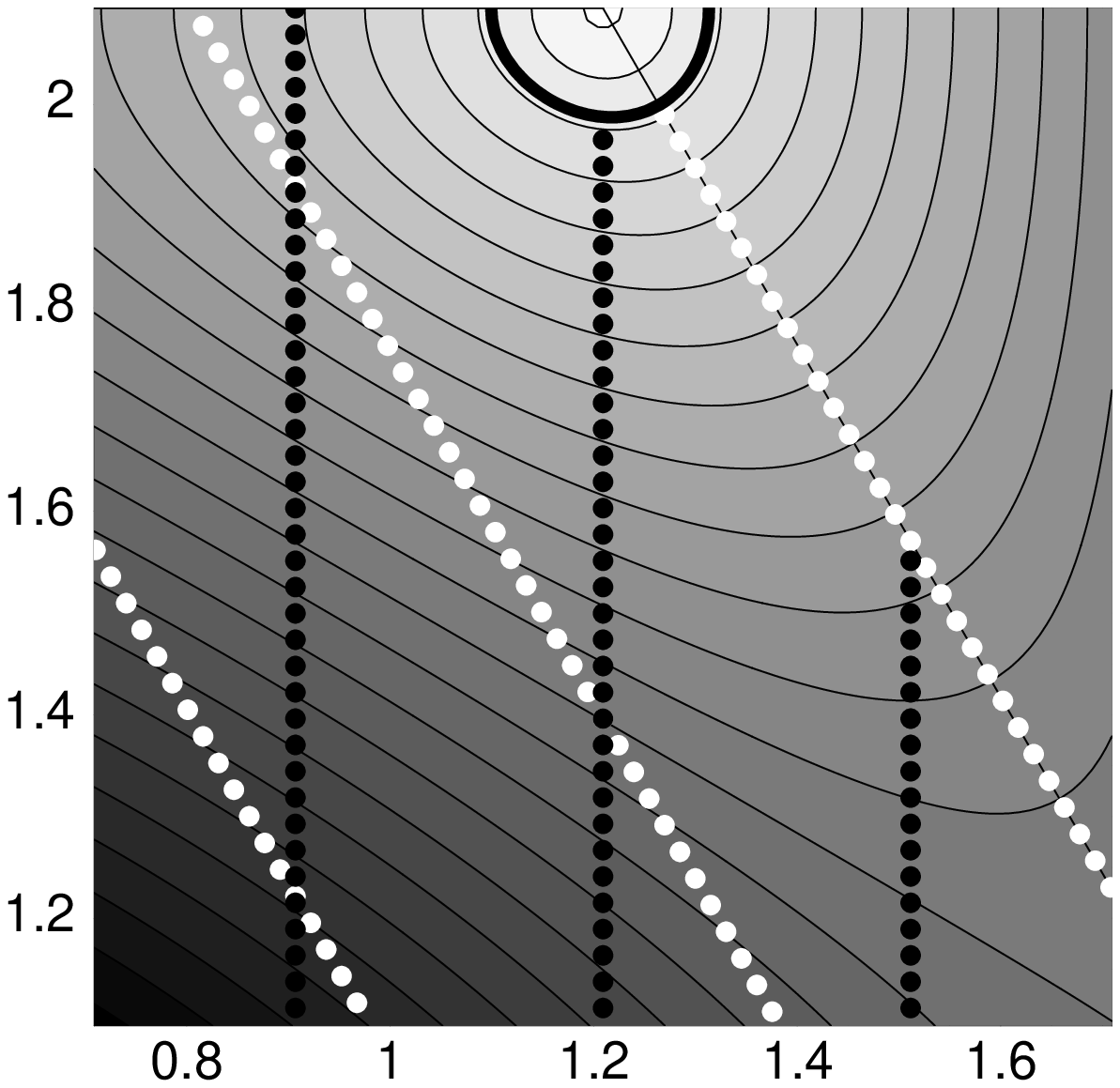}\\
\vspace{0.5cm}
\includegraphics[width=4cm]{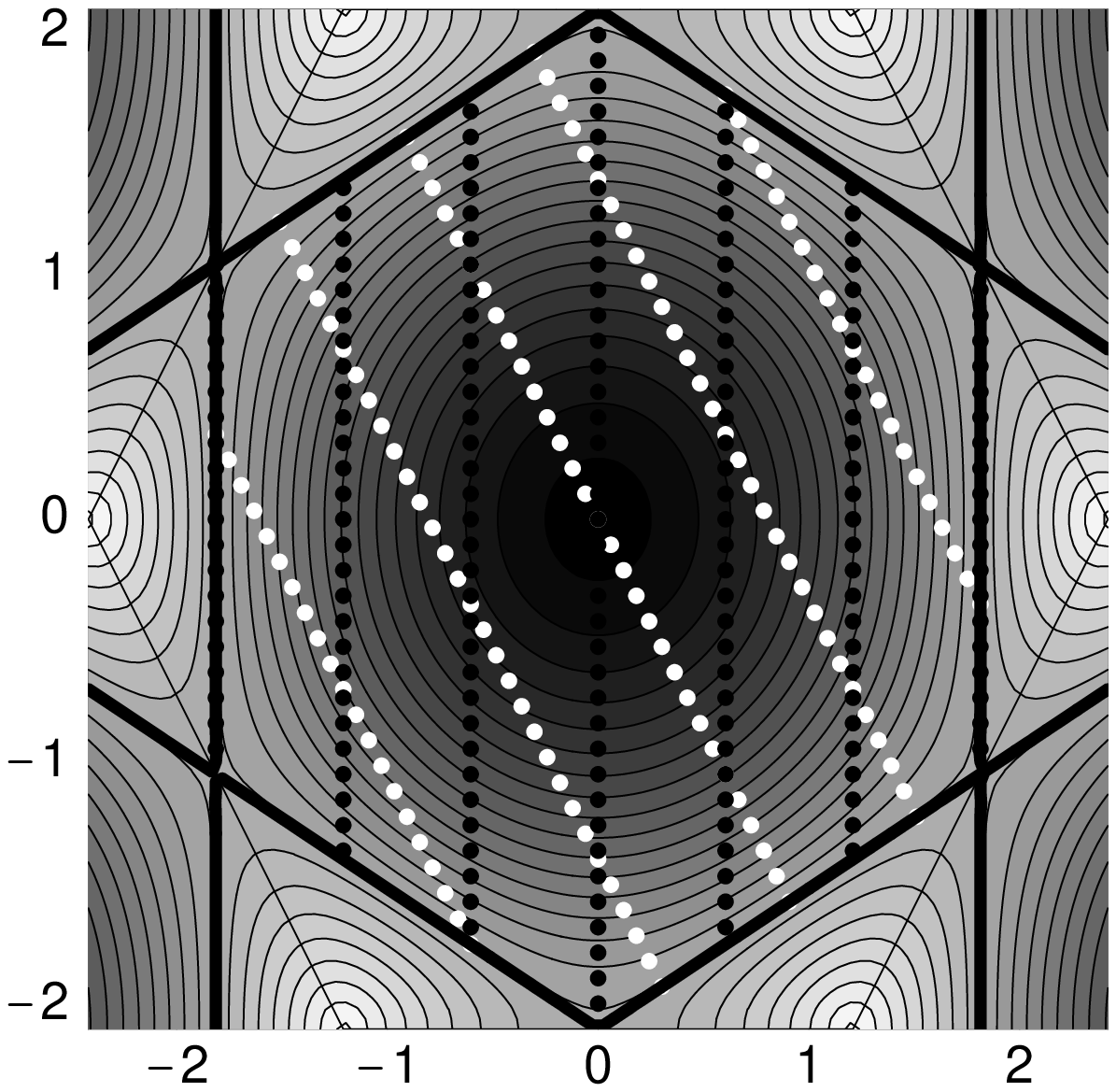}\hspace{0.5cm}
\includegraphics[width=4cm]{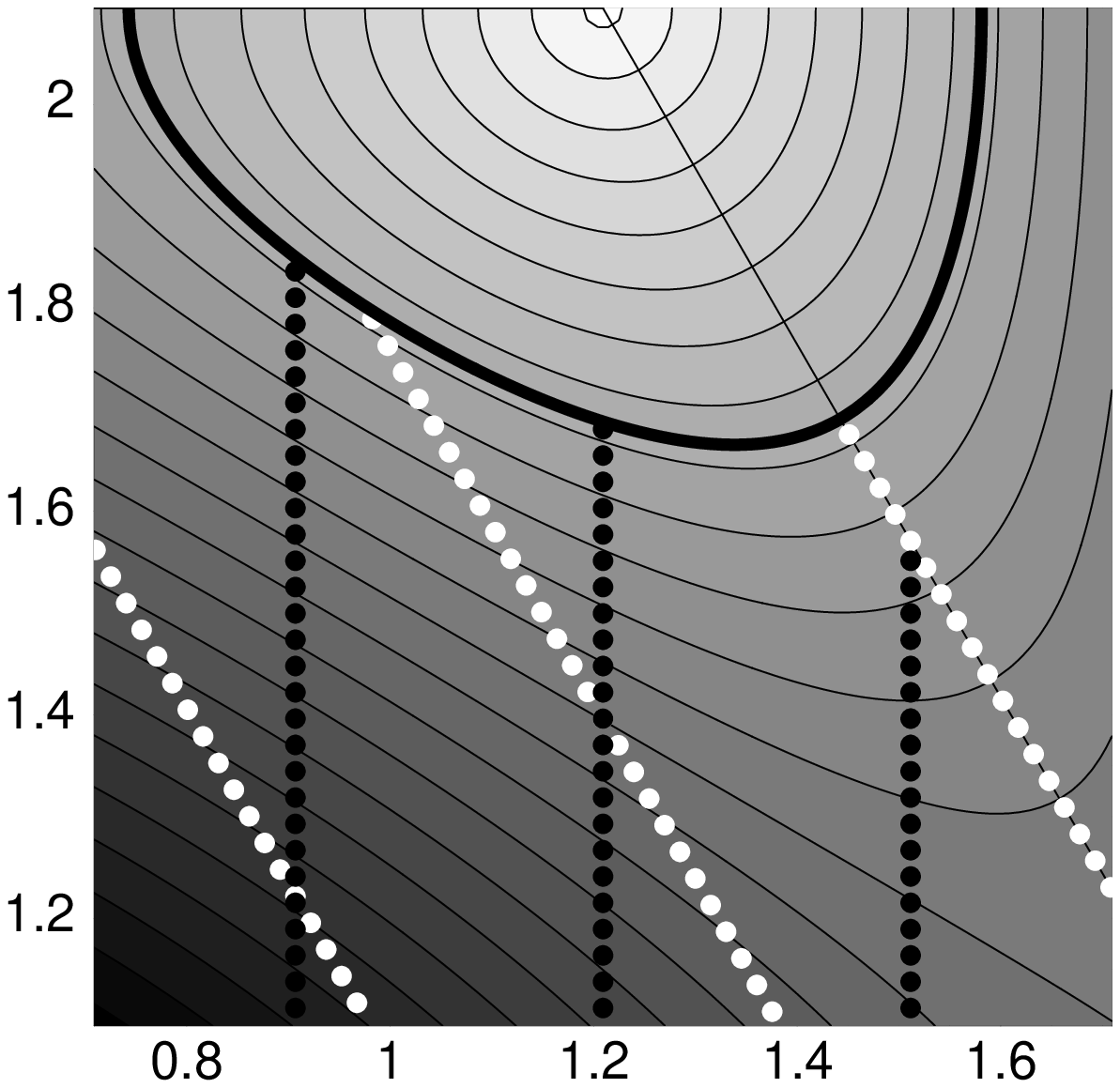}\\
\begin{picture}(0,0)
\put(-140,240){\bf \large a)}
\put(0,240){\bf \large b)}
\put(0,120){\bf \large c)}
\put(-140,120){\bf \large d)}
\end{picture}
\caption{\label{fig:bzones} The Brillouin zone and momentum states of an  arbitrary armchair (white dots)
and zigzag (black dots) nanotube of similar radius, at $\phi = 0$. The background
is the contour plot of $E_{\mathbf k}$ from Eq. (\ref{dispersion}). The Fermi contour is marked by thick
solid lines. a) Full Brillouin zone, undoped nanotube ($\mu_{chem} = 0$). The Fermi surface
reduces to two points at the vertices of the hexagon. b) The neighbourhood of the Fermi
point $(2\pi/(3\sqrt{3}), 2\pi/3$), $\mu_{chem} = -0.16\gamma$. The contour is circular and 
both nanotubes respond with almost the same magnetic moments (see Fig. \ref{fig:smallcone-constN}), momentum lines cross the Fermi contour
at identical angles. c) The neighbourhood of the Fermi point $(2\pi/(3\sqrt{3}), 2\pi/3)$, 
$\mu_{chem} = -0.6\gamma$. The Fermi contour loses the rotational symmetry and magnetic moments
in zigzag and armchair nanotubes differ (see Fig. \ref{fig:carpets}). d) Full Brillouin zone, $\mu_{chem} = -\gamma$.
The Fermi contour is a closed hexagon, with two sides parallel to the momentum lines in the
zigzag nanotube. Currents in this tube are very strongly enhanced, in others nearly vanishing. }
\end{figure}
%
%
%
{\bf Isolated or connected: $\mu_{chem} = const$ versus $N_e = const$ approach. }
Different results were obtained for $\mu_{chem} = const$ (nanotube in a circuit) and $N_e = const$ (isolated nanotube). Both conditions can be realized experimentally.  
The calculations were performed for both electron and hole doping, changing the Fermi level from $E_F = 0$
(no doping) to $E_F = \pm\gamma$. In the case of nanotubes
doped to $E_F = \pm \gamma$ we found that again there is no difference between 
$N_e = const$ and $\mu_{chem} = const$, because the Fermi contour has then the hexagonal
symmetry of the Brillouin zone (cf. Fig. \ref{fig:bzones}d) and the number of momentum states within it is constant with $\phi$. At intermediate doping the two approaches give distinctly different results
which are shown in Fig. \ref{fig:smallcone-both}.\\
%
%
%
%
%
{\bf Dependence of $\mu_{orb} (\phi)$ on the nanotube radius.}
The independence of $\mu(\phi,T)$ on the radius of the nanotube shown in Fig. \ref{fig:irsquare}
is characteristic for undoped nanotubes only. The sum in Eq. (\ref{muorb-with-T}) is in that case over the whole
Brillouin zone and the dominating currents come from the momentum states in the neighbourhood 
of the Fermi points $\mathbf{K}_i$, where the dispersion relation is conical. On the other hand, if the doping is to $E_F = \pm \gamma$,
the amplitude of $\mu_{orb}(\phi,T)$ grows linearly with the radius of the nanotube 
(for illustration in zigzag case, see Fig. \ref{fig:irsquare}). The currents dominating the sum are those from the states near the Fermi contour which in this case is the inner
hexagone in Fig. \ref{fig:bzones}d; the dispersion relation
below $-\gamma$ is nearly parabolic.
As we lower the chemical potential of the nanotube, we cross from the regime of the
magnetic moment independent of $R$ (conical dispersion relation near the Fermi points) to the regime where
it depends linearly on $R$ (quasi parabolic $E(\mathbf{k})$). The Brillouin zone 
in the intermediate regime 
is shown in Figs. \ref{fig:plane-holes} and \ref{fig:bzones}b \& c. The
magnetic moment depends on the nanotube radius in a more complicated way (an example
is shown in Fig. \ref{fig:irsquare} as open circles).
Note that its radius dependence in the case of $E_F = \pm \gamma$
resembles that of persistent currents in metallic cylinders \cite{ibm,steb}.\\
{\bf Electron-hole symmetry of $\mu_{orb}(\phi,T)$.}
In temperatures up to 300K the $\mu_{orb}(\phi,T)$ for electron and hole doping are identical.
This feature is found both when  $\mu_{chem} = const$ and when $N_e = const$ 
(see Figs. \ref{fig:armch}--\ref{fig:zigzag}, also Figs. \ref{fig:smallcone-both} and
\ref{fig:smallcone-constN}). It is a combined effect of the symmetry 
of the Fermi-Dirac function with respect to the chemical potential, and the hole-electron 
symmetry of the dispersion relation of graphene which both enter Eq. (\ref{muorb-with-T}) 
(for experimental confirmation see \cite{kouwenhoven}).\\
{\bf Dependence of $\mu_{orb}(\phi, T)$ on the chirality of the nanotube.}
The shape of the Fermi contour in a CN changes significantly with doping \cite{PhyLe},
from two points (Fig. \ref{fig:bzones}a), through a set of increasingly flattened circles (Fig. \ref{fig:bzones}b,c) , to a hexagone (Fig. \ref{fig:bzones}d).
That is why in doped nanotubes the form of the magnetic moment as a function of magnetic field
and temperature depends strongly on the chirality of the nanotube and reflects the
geometrical relation of their momentum lines to the actual shape of the Fermi surface.
The paramagnetic contribution to the persistent current and consequently to the magnetic moment is enhanced if the number of states
crossing simultaneously the Fermi level is large, therefore the strongest
magnetic response should be achieved in systems with momentum lines
nearly parallel to the Fermi surface \cite{PhyLe}. This is achieved in a zigzag CN doped to
$E_F = \pm\gamma$, where the lines of states (marked by black dots in Fig. \ref{fig:bzones}d) 
are parallel to two sides of the Fermi surface (inner hexagone). The
resulting magnetic moment is huge -- see Fig. \ref{fig:zigzag} and 
Tables \ref{tab:constM} and \ref{tab:constN}. \\
The shape of $\mu_{orb}(\phi, T)$ is different in isolated and connected nanotubes.
This difference is illustrated in Figs. \ref{fig:armch} - \ref{fig:zigzag} for fields accessible in labs 
(up to 40T), and in Fig. \ref{fig:smallcone-both} for the full $\phi_0$ period.

%
%
%
%

Note for instance that in an isolated nanotube ($N_e = const$) the magnetic moment
at half-filling (the line in the centre of the bottom row in Fig. \ref{fig:smallcone-both}) changes completely when even a few electrons are added or removed from the system. At constant chemical potential the dependence of $\mu_{orb}(\phi, T)$
on doping is more smooth.\\
{\bf Temperature dependence of $\mu_{orb}(\phi,T)$.} 
The main effects of temperature on $\mu_{orb}(\phi, T)$ are a suppression of its amplitude 
and its smoothing into a sinusoidal shape. The former effect causes a decrease of the
magnetic moment by a factor 2-3 over the temperature range from 0 to 300 K.  Thus the measurements 
could be performed at room temperature which simplifies considerably the experiments.\\
The latter effect smoothes out all sharp features in the $\mu_{orb}(\phi, T)$ dependence
and can even turn a diamagnetic moment at $T = 0$ into a paramagnetic one
at $T = 300 K$. This effect is seen for both isolated and connected nanotubes (see Fig. \ref{fig:armch})\\

The sawtooth shape of the orbital magnetic moment in a $(38,36)\times(-4028,4101)$ nanotube
doped to $\pm\gamma$ in Fig.\ref{fig:armch} is a manifestation of the fractional period of the
Aharonov-Bohm effect, noticed by Sasaki et al. \cite{sasaki}. In CNs
whose length is commensurate with their circumference, the period of the
Aharonov-Bohm oscillations is a fraction of $\phi_0$.

The full dependence of $\mu_{orb}(\phi)$ on the doping, both in isolated 
($N_e = const$) and connected ($\mu_{chem} = const$) nanotubes shows several interesting 
features.\\
First, as long as the doping does not exceed $\pm 0.3\gamma$, the orbital magnetic moments
fall in two distinct classes, depending on whether the nanotube was originally metallic
or semiconducting. This feature appears in both conditions, in Fig. {\ref{fig:smallcone-constN}} we
show it for $N_e = const$. It is due to the fact that the Fermi surface for small dopings
is a set of nearly circular loops, meeting at the same angle the momentum lines of any chirality
(cf. Fig. \ref{fig:bzones}b).\\

%
%
\begin{figure*}
\includegraphics[width=14cm]{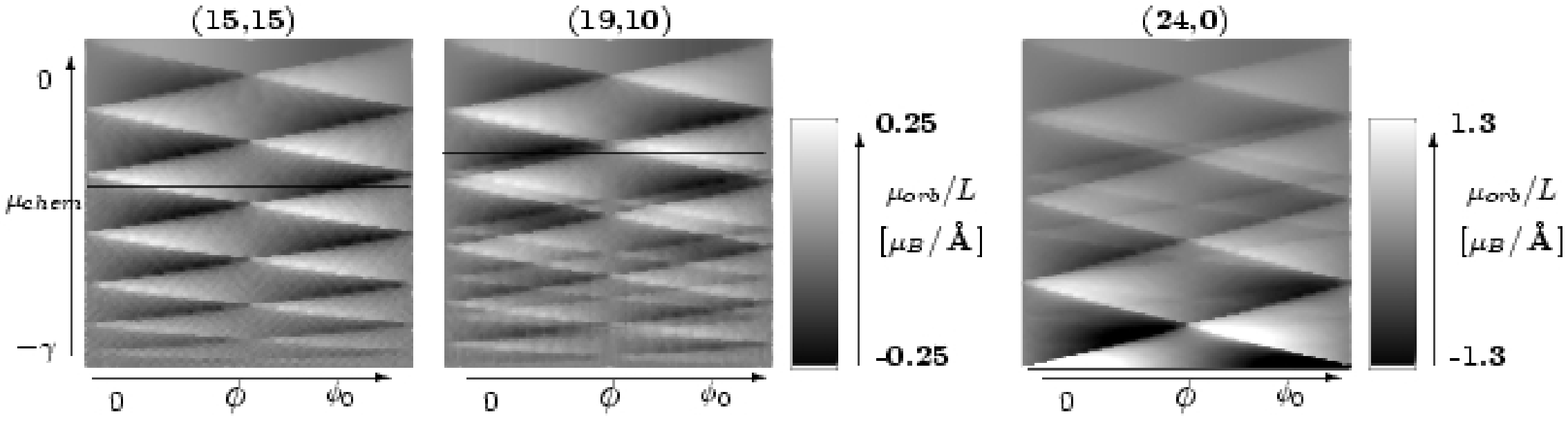}\\
\caption{\label{fig:carpets} The magnetic moment per {\AA} (colour scale) in  (15,15) armchair (left),
(19,10) chiral metallic (middle) and (24,0) metallic zigzag (right) hole-doped nanotubes
($R = 10.2$\AA). The range of chemical potential is from $-\gamma$ to 0. 
The part corresponding to electron-doped nanotube
is symmetrical to the hole-doped. At the chemical potential lower than $\sim -0.3\gamma$
(lower part of the plots), the Fermi contour ceases to be circular and the 
$\mu_{orb}$ patterns become different for different chiralities. The maximum amplitude
of the magnetic moment is in each nanotube reached at a different value of doping, marked by
a straight line.}
\end{figure*}

%
%
As the Fermi surface loses the circular symmetry, at larger values of doping, differences
appear between nanotubes of the same type of conduction and different chiralities (cf. Fig. \ref{fig:bzones}c). 
This effect shows both in the case of nanotube isolated and forming part of a circuit. 
In Fig. \ref{fig:carpets} we show it for $\mu_{chem} = const$. The doping level
at which the maximum amplitude of the magnetic moments is reached (marked by straight lines
in Fig. \ref{fig:carpets}) depends on the chirality of the CN.\\
%
%
%
%
%
\section{Conclusions}

CNs can be one of the basic ingredients of the future nanoelectronic devices. Therefore
their behaviour in the magnetic field is of big importance.
The magnetic field has a strong effect on the electronic structure of the CN -- it can be used to tune
the energy spectrum. This raises the possibility of controlling the energy levels through an 
external field, opening the door to further studies of fundamental properties of nanotubes
as well as technological applications.\\
When applied parallel to the tube axis, the magnetic field creates orbital magnetic moments,
which depend strongly on the chirality, length and doping level of the CN.\\

The results of our model calculations show that the behaviour of the orbital magnetic 
moment in single-wall nanotubes has the following features:\\
1) the temperature diminishes its amplitude and smoothes out sharp features in 
the $\mu_{orb}(\phi,T)$ dependence, which in doped nanotubes can even change the character 
of the response from diamagnetic
at low T to paramagnetic at high T (cf. Fig. \ref{fig:armch}).\\
2) its dependence on $\phi$ is the same in electron- and hole-doped CNs in temperatures up to 300K,
in nanotubes both isolated and forming part of a circuit\\
3) is nearly independent of $R$ in half-filled CNs and depends on $R$ in the doped ones\\
4) scales with length in both pure and doped CNs\\
5) is independent of chirality for pure CNs and depends strongly on chirality in doped nanotubes.\\

It is well known that in the magnetic field parallel to the axis a metallic CN can be converted
into a semiconducting one and vice versa \cite{ajiki1}. This has been recently demonstrated
as a change of the band gap in a series of experiments \cite{mceuen,bezryad,smalley}.
This extraodrinary feature can be clearly seen also in the $\phi$ dependence of the orbital
magnetic moment.\\
The nanotubes for which $\mu_{orb}(\phi, T)$ exhibits a steep rise are metallic, otherwise they
are semiconducting. Inspection of  e.g. Fig. \ref{fig:chirality-independent} shows
that the nanotube denoted by (m), metallic at $\phi \simeq 0$ transforms into a semiconducting
one at $\phi \neq n\phi_0$ ($n \in \mathbb{Z}$), whereas the semiconducting nanotube denoted
by (s) becomes metallic at $\phi = 1/3\; \phi_0$ and $\phi = 2/3\; \phi_0$.  The maximum values of $\mu_{orb}$ in the range of 0-20 T are given in Tables
\ref{tab:constM} and \ref{tab:constN}.\\

The relation between $\mu^F_{orb}$ calculated and measured
in \cite{mceuen} and $\mu_{orb}(\phi, T)$ discussed in this paper is the following.
$\mu^F_{orb}$, being the magnetic moment of an electron at (or close to) the Fermi point, is calculated
in the energy range where its dispersion relation is a linear function of $\mathbf{k}$, therefore
$\mu^F_{orb}$ is independent of the flux. The total magnetic moment $\mu_{orb}(\phi, T)$ is
calculated as a sum over all $\mathbf{k}$ states, thus containing terms with nonlinear dispersion
relation which cause the flux dependence of the magnetic moment. Moreover, $\mu_{orb}(\phi, T)$ 
calculated from Eq. (\ref{muorb-with-T}) takes into account the temperature dependence of the
energy level occupation. This effect has been neglected in the
calculation of  $\mu^F_{orb}$.\\
Whereas the measurements of the band gap as a function of $B$ give the information about
 $\mu^F_{orb}$ (which is actually the orbital moment unit, $\mu^F_{orb} = \pi R^2 I_0$, where 
$I_0 = \frac{e v_F}{2\pi R}$ is the unit of the persistent current), the full   $\mu_{orb}(\phi,T)$ 
could be measured in other experiments which will be sensitive to it.\\
In our opinion
this measurement may be possible in a setup with a double walled CN. The outer shell
can be used as a field detector. The current running along it would be modified by the 
Aharonov-Bohm effect. The precise measurement of this current, compared with the current
in identical but single-walled CN could reveal the magnetic moment produced by the inner tube.\\
\begin{table*}
\begin{tabular}{|c|c|c|c|c|}
\hline 
  &  $E_F = 0$  &  $E_F = \pm0.3 \gamma$  &  $E_F = \pm0.6 \gamma$  & $E_F = \pm\gamma$ \\
\hline 
\hline
 armchair (37,37)  & 78.7 {\em (42.9)} & {\bf -100} {\em(-78.9)} &  -99.5 {\em(-65.1)} & $\leq$ 3 ($\leq$ {\em 1}) \\
\hline
 chiral S (38,36)  & -28.2{\em (-20.5)} & 42.6 {\em(43.5)} & 93.2 {\em(48.6)} & -26.4 ( $\leq$ {\em 1}) \\
 chiral M (48,24)  & 78.7 {\em(43.3)} & -90.6 {\em(63.9)} & {\bf 161.6} {\em(66.4)} & -9.7 ( $\leq$ {\em 2}) \\
 chiral S (49,23)  & -27.5 {\em(-20)} & {\bf 218.8} {\em(96.7)} & -89.1 {\em(-46.3)} & $\leq$ 2 ( $\leq$ {\em 2}) \\
 chiral S (63,2)  & -29.3{\em (-21.8)} & {\bf -116.9} {\em(-72.7)} & {\bf -148.1} {\em(31.1)} & {\bf 399.2 {\em(389)}} \\
\hline
 zigzag M (63,0)  &  80.2 {\em(44.1)} & -62.5 {\em(-28.7)} & 28.7 {\em(-63.1)} & {\bf -714.7 {\em(-523)}} \\
 zigzag S (64,0)  & -29.8 {\em(-21.9)} & {\bf -113.3} {\em(-73.2)} & {\bf -290} {\em (21.8)} & {\bf 3559}, at $B \ll 1$T \\
  & & & & {\bf {\em(818.2)}} \\ 
\hline  
\end{tabular}
\caption{\label{tab:constM}Magnetic orbital moment at $\mathbf{\mu_{chem} = const } $  for SWNT of radius 25\AA, length 0.1$\mu$m, for various
chiralities and levels of electron or hole doping $E_F$. The unit is $\mu_B$, values in brackets (italics) are
for $T = 300$K, outside brackets  for $T = 0$K. The orbital moments greater than $100\mu_B$ 
are in bold font. The values in the table are the maximum
values obtainable in magnetic fields ranging from 0 to 20T. Typically this maximum value
is reached at 20T, except for one case, (64,0) doped to $\pm\gamma$.}
\end{table*}
\begin{table*}
\begin{tabular}{|c|c|c|c|c|}
\hline 
  &  $E_F = 0$  &  $E_F = \pm0.3 \gamma$  &  $E_F = \pm0.6 \gamma$  & $E_F = \pm\gamma$ \\
\hline 
\hline
 armchair (37,37)  & 78.7 {\em } & {\bf -107} {\em } & {\bf -105.5}  {\em } & 4.44  \\
\hline
 chiral S (38,36)  & -28.2{\em } &  -41.9 ($B\simeq 12$T) {\em } &  77.7 {\em } & 19.6 ($B\simeq 18$T)  \\
 chiral M (48,24)  & 78.7 {\em } &  -92.7 {\em } & {\bf 194.5} ($B\simeq 14$T) {\em } & -9.65  \\
 chiral S (49,23)  & -27.5 {\em } & {\bf 225 } {\em } &  -94.8 {\em } & -3.48  \\
 chiral S (63,2)  & -29.3{\em } & {\bf -115.4} {\em } & {\bf -223.8} {\em } & {\bf 395.9 {\em }} \\
\hline
 zigzag M (63,0)  &  80.2 {\em } &  -69.5 {\em } & -11.9 {\em } & {\bf -714.7 {\em }} \\
 zigzag S (64,0)  & -29.8 {\em } & {\bf } {\em } & {\bf -293} {\em  } & {\bf 3559}, at $B \ll 1$T \\
  & & & & {\bf {\em}} \\ 
\hline  
\end{tabular}
\caption{\label{tab:constN}Magnetic orbital moment at $\mathbf{N_e = const } $  for SWNT of radius 25\AA, length 0.1$\mu$m, for various
chiralities and levels of electron or hole doping. The unit is $\mu_B$,  $T = 0$K. The orbital moments greater than $100\mu_B$ 
are in bold font. The values in the table are the maximum
values obtainable in magnetic fields ranging from 0 to 20T. Typically this maximum value
is reached at 20T, if  otherwise, the corresponding approximate value of $B$ is indicated.}
\end{table*}
In this paper we have given a survey of the properties of the orbital magnetic moments
for single-wall nanotubes of different chiralities with different radii and length, and for various
values of electron or hole doping, at zero and finite temperatures. In multiwall nanotubes
these dependences are more complex, since all shells are penetrated by the magnetic
field and all must be taken into account. The magnetic moments are 
then a superposition of moments from different shells. Some aspects of this problem have 
recently been discussed in \cite{actapol,physrev}.

\begin{acknowledgments}
This work was supported by the  Polish
Ministry of Scientific Research and Information Technology under the
(solicited) grant No PBZ-MIN-008/P03/2003.\\
M.S. thanks  H\'{e}l\`{e}ne Bouchiat for valuable discussions.
\end{acknowledgments}

\end{document}